\documentclass{PoS}
\usepackage{color}
\usepackage{amsmath}

\title{A high-statistics study of the nucleon EM form factors, axial charge and quark momentum fraction}

\ShortTitle{Nucleon EM form factors, axial charge and quark momentum fraction}

\author{ \speaker{\mbox{B.~J\"ager\thanks{Speaker.}}, T.D.~Rae}, S.~Capitani, M.~Della~Morte, D.~Djukanovic, G.~von~Hippel,
	B.~Knippschild, H.B.~Meyer, H.~Wittig\\
       
     PRISMA Cluster of Excellence and Institut f\"ur Kernphysik, Becher-Weg 45, University of
	Mainz, D-55099 Mainz, Germany\\
	Helmholtz Institute Mainz, University of Mainz, D-55099 Mainz, Germany\\
	IFIC and CSIC, calle Catedratico Jose Beltran 2, 46980 Paterna, Spain\\
 	E-mail: \email{jaeger@kph.uni-mainz.de, thrae@uni-mainz.de}}

\abstract{

We present updated results for the nucleon axial charge and electromagnetic (EM) form factors, 
which include a significant increase in statistics for all ensembles (up to 4000 measurements), 
as well as the addition of ensembles with pion masses down to $m_\pi\sim195$ MeV. 
We also present results for the average quark momentum fraction.  
The new data allows us to perform a thorough 
study of the systematic effects encountered in the lattice extraction. We concentrate on systematic effects
due to excited-state contaminations for each of the quantities, which we check using several 
different time separations between the operators at the source and sink 
through a comparison of plateau fits and the summed operator insertion method 
(which provides a mechanism to suppress the excited-state contamination).
We confirm our earlier finding \cite{Capitani:2012gj} that a reliable extraction of the axial
charge must be based on a method which eliminates excited-state contaminations. 
Similar conclusions apply to our EM form factor calculations \cite{Capitani:2012ef}.
The measurements are calculated using the CLS ensembles with 
non-perturbatively O(a) improved Wilson fermions in $N_f=2$ QCD.\newline

\begin{flushright}
\includegraphics[width=0.28\linewidth]{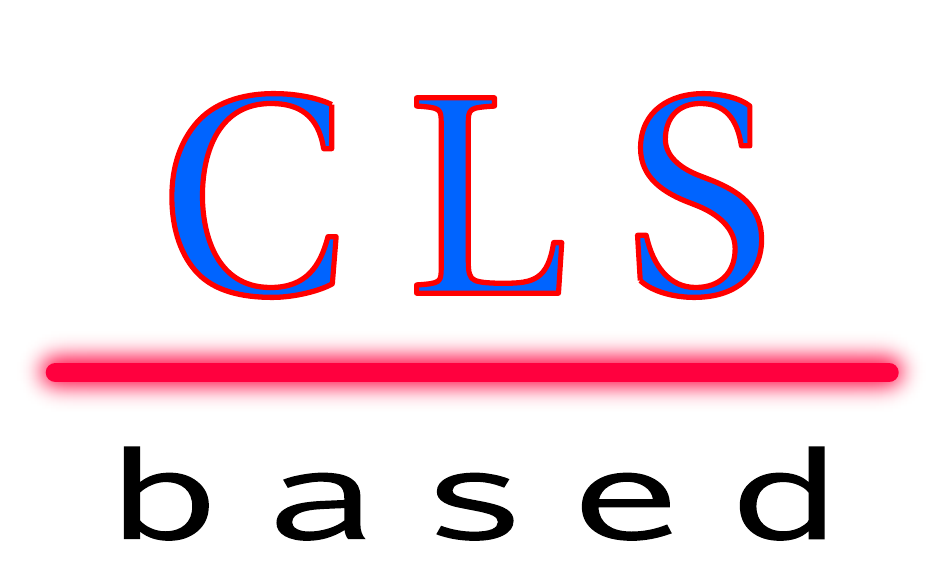}
\end{flushright}
}

\FullConference{31st International Symposium on Lattice Field Theory LATTICE 2013\\
                 July 29 - August 3, 2013\\
                 Mainz, Germany}

\begin{document}

\section{Introduction\label{seci}}
\noindent
The axial charge of the nucleon is very well determined from experiment, $g_A=1.2701(25)$ \cite{Beringer}, 
and provides a benchmark for Lattice QCD (LQCD) calculations, since it is constructed as a 
simple matrix element from a local operator with quark bilinears, involves no momentum in
the initial and final states, and is an isovector quantity that has no quark-disconnected 
diagrams. However, the results from lattice calculations are typically $\sim10\%$ below the experimental value 
\cite{Alexandrou:2010cm}-\cite{syritsyn:2013}.
It is therefore important to ensure that systematic effects are under sufficient control.
We have previously argued in \cite{Capitani:2012gj} that this discrepancy can be explained by 
carefully accounting for excited states, for which we use the summation method
(described in section~\ref{seciii}). Similar methods have been used in \cite{Green:2012ud}. This 
proceedings contribution provides an update to the results for our $g_A$ calculation \cite{Capitani:2012gj}, and to our nucleon electromagnetic 
(EM) form factors results \cite{Capitani:2012ef,Capitani:2010sg,Capitani:2012uca}. The EM form 
factors are crucial observables in hadronic physics and provide details of the distribution 
of charge and magnetisation in the nucleon, for which a similar discrepancy between the lattice and experiment
is seen as for $g_A$ \cite{Alexandrou:2010cm,Renner:2010ks},\cite{nuclFF:RBC08_nf2}-\cite{syritsyn:2013},\cite{Green:2012ud,Collins:2011mk}.
In addition to
the quantities previously calculated by our group, we present first results for the quark momentum fraction $\langle x \rangle$
of the nucleon, which may also be considered a benchmark quantity and tends to be 
overestimated in lattice calculations \cite{nuclFF:LHPC10_nf2p1,nuclFF:QCDSF_lat10,syritsyn:2013,Green:2012ud},
\cite{Syritsyn:2009mx}-\cite{Aoki:2010xg}.
Our simulations use non-perturbatively $\mathcal{O}(a)$ 
improved Wilson fermions in $N_f=2$ QCD, generated as part of the CLS effort. Table~\ref{ensembles} 
provides details of the lattice ensembles.

\begin{table}
\begin{center}
\begin{tabular}{cccccccc}
    	\hline
    	$\beta$ & $a$ $[\mathrm{fm}]$ & lattice & $L$ $[\mathrm{fm}]$ & 
    	$m_\pi$ $[\mathrm{MeV}]$ & $m_\pi L$ & Label & \# meas.\\
    	\hline
    	$5.20$ & $0.079$ & $64 \times 32^3$ & $2.5$ & $473$& $6.0$& A3 & 2128\\
    	$5.20$ & $0.079$ & $64 \times 32^3$ & $2.5$ & $363$& $4.7$& A4 & 3200\\
    	$5.20$ & $0.079$ & $64 \times 32^3$ & $2.5$ & $312$& $4.0$& A5 & 4000\\
    	$5.20$ & $0.079$ & $96 \times 48^3$ & $3.8$ & $262$& $5.0$& B6 & 2544\\
    	\hline
    	$5.30$ & $0.063$ & $64 \times 32^3$ & $2.0$ & $451$& $4.7$& E5 & 4000\\
    	$5.30$ & $0.063$ & $96 \times 48^3$ & $3.0$ & $324$& $5.0$& F6 & 3600\\
    	$5.30$ & $0.063$ & $96 \times 48^3$ & $3.0$ & $277$& $4.2$& F7 & 3000\\
    	$5.30$ & $0.063$ & $128 \times 64^3$ & $4.0$ & $195$& $4.0$& G8 & 4176\\
    	\hline
	$5.50$ & $0.050$ & $96 \times 48^3$ & $2.4$ & $536$& $6.5$& N4 & 600\\
    	$5.50$ & $0.050$ & $96 \times 48^3$ & $2.4$ & $430$& $5.2$& N5 & 1908\\
	$5.50$ & $0.050$ & $96 \times 48^3$ & $2.4$ & $340$& $4.0$& N6 & 3784\\
    	$5.50$ & $0.050$ & $128 \times 64^3$ &$3.2$ & $270$ & $4.4$& O7 & 1960\\
    	\hline    	
\end{tabular}
\caption{\label{ensembles} Details of the lattice ensembles used in this study, showing $\beta$-values, 
lattice spacing $a$ (determined in \cite{Capitani:2011fg}), lattice extent $L$ (where $T=2L$), pion mass $m_\pi$ and the total number of measurements.}
\end{center}
\end{table}
~\newline
The matrix element of a nucleon interacting with the axial current, $A_\mu=\overline{\psi}(x)\gamma_5\gamma^\mu \psi(x)$, 
may be decomposed into the axial and pseudoscalar form factors $G_A$ and $G_P$:
\begin{equation}
\langle N(p^\prime,s^\prime)|A_\mu|N(p,s)\rangle=\bar{u}(p^\prime,s^\prime)\left[\gamma_\mu \gamma_5 G_A(Q^2)+\gamma_5\frac{q_\mu}{2m_N}G_P(Q^2)\right]u(p,s),
\end{equation}
whereas for the electromagnetic current, $V^\mu=\overline{\psi}(x)\gamma^\mu \psi(x)$, the matrix element 
may be parameterised by the Dirac and Pauli form factors $F_1$ and $F_2$:
\begin{equation}
\langle N(p^\prime,s^\prime)|V_\mu|N(p,s)\rangle=\bar{u}(p^\prime,s^\prime)\left[\gamma_\mu F_1(Q^2)+i\frac{\sigma_{\mu\nu} q_\nu}{2m_N}F_2(Q^2)\right]u(p,s),
\end{equation}
where $u(p,s)$ is a Dirac spinor with spin $s$, and momentum $p$, $\gamma_\mu$ is a Dirac matrix, 
$\sigma_{\mu\nu}=\frac{1}{2i}[\gamma_\mu,\gamma_\nu]$, and $Q^2=-(E_{p^\prime}-E_p)^2+\vec{q}^2$ 
where $\vec{q}=\vec{p}^\prime-\vec{p}$. The Pauli and Dirac form factors are related to the Sachs form factors 
$G_E$ and $G_M$, 
\begin{equation}
G_E(Q^2)=F_1(Q^2)-\frac{Q^2}{4m_N^2}F_2(Q^2), \qquad G_M(Q^2)=F_1(Q^2)+F_2(Q^2),
\end{equation}
that are measured in scattering experiments via the differential cross section described by the Rosenbluth formula.
The form factors may be Taylor expanded in the momentum transfer $Q^2$,
\begin{equation}
G_X(Q^2)=G_X(0)\left(1-\frac{1}{6}\langle r_X^2 \rangle Q^2+\mathcal{O}(Q^4)\right),
\end{equation}
from which the charge radii of the nucleon may be determined:
\begin{equation}
\langle r_X^2 \rangle = -\frac{6}{G_X(Q^2)}\frac{\partial G_X(Q^2)}{\partial Q^2}\Bigg|_{Q=0},
\end{equation}					
where $X=E,~M$. Note that $G_A(0)=g_A$ and for the conserved current, $G_E(0)=1$ and $G_M(0)=\mu$, where $\mu$ measures 
the magnetic moment in nuclear magneton units $e/(2m_N)$.
~\newline

\noindent
The hadronic matrix element containing a single derivative can be related to the generalised form 
factors $A_{20}$, $B_{20}$ and $C_{20}$ through
\begin{multline}
\langle N(p^\prime,s^\prime)|\gamma_{\{\mu}\stackrel{\leftrightarrow}{D}_{\nu \} }|N(p,s)\rangle= \bar{u}(p^\prime,s^\prime)\Big(\gamma_{\{ \mu}Q_{\nu \} } A_{20}(Q^2)\\
+i\frac{\sigma_{\{ \mu\alpha}Q_\alpha p_{\nu \} }}{2m}B_{20}(q^2)+\frac{1}{m}p_{ \{ \mu }p_{\nu \} } C_{20}(Q^2)\Big)u(p,s),
\end{multline}
where $\stackrel{\leftrightarrow}{D}_{\nu}=\stackrel{\rightarrow}{D}_{\nu}-\stackrel{\leftarrow}{D}_{\nu}$, 
and $A_{20}(0)\equiv\langle x \rangle$ is the average quark momentum fraction.

\section{Lattice formulation\label{secii}}
\begin{figure}
\centering
\includegraphics[width=0.45\linewidth]{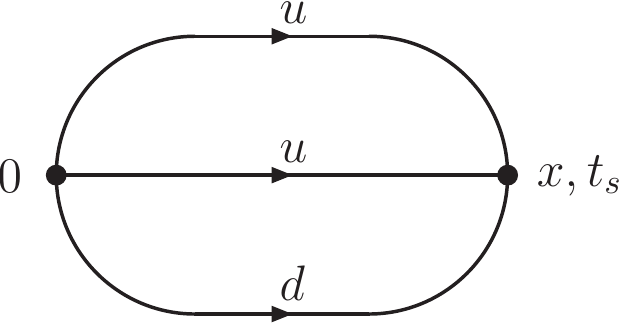}
\hspace{1cm}
\includegraphics[width=0.45\linewidth]{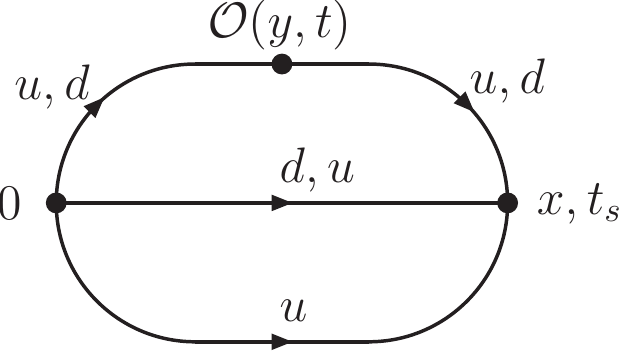}\\
\caption{\label{schem} Schematic diagrams for the two- and three-point functions, left and right panels respectively.}
\end{figure}
\noindent
The calculation of the form factors requires a ratio of correlation functions, for which we use 
\begin{equation}
R_{\gamma_\mu}(\vec{q},t,t_s)=\frac{C_{3,\gamma_\mu}(\vec{q},t,t_s)}{C_2(\vec{0},t_s)}\sqrt{\frac{C_2(\vec{q},t_s-t)C_2(\vec{0},t)C_2(\vec{0},t_s)}{C_2(\vec{0},t_s-t)C_2(\vec{q},t)C_2(\vec{q},t_s)}},
\label{Ratio}
\end{equation}
where $\vec{p}^\prime=0$. This ratio was found to be the most effective ratio of several studied 
in \cite{Alexandrou:2008rp}. In the case $\vec{p}=\vec{q}=0$, which is relevant for $g_A$ and $\langle x \rangle$, eq.~\ref{Ratio},
simplifies as the square root factor equals 1.  
The two- and three-point functions $C_2(\vec{p},t)$ and $C_{3,\gamma_\mu}(\vec{q},t,t_s)$ are given by (fig.~\ref{schem}),
\begin{eqnarray}
C_2(\vec{p},t)&=&\sum_{\vec{x}}\langle\Gamma_{\alpha^\prime\alpha}J_\alpha(x)\overline{J}_{\alpha^\prime}(0)\rangle e^{-i\vec{p}.\vec{x}},\\
C_{3,\gamma_\mu}(\vec{q},t,t_s)&=&\sum_{\vec{x},\vec{y}}\langle\Gamma_{\alpha^\prime\alpha}J_\alpha(\vec{x},t_s)\mathcal{O}_{\gamma_\mu}(\vec{y},t)\overline{J}_{\alpha^\prime}(0)\rangle e^{-i\vec{q}.\vec{y}},
\end{eqnarray}
where $J_\alpha(x)$ is a suitably chosen interpolating 
operator with the correct quantum numbers to create a nucleon, and $\Gamma_{\alpha\alpha^\prime}$ 
is a projection matrix used to give the interpolating fields the correct parity. We chose to polarise 
the nucleon in the $z$-direction,  $\Gamma=\frac{1}{2}(1+\gamma_0)(1+i\gamma_5\gamma_3)$.  
We consider both local and conserved vector 
currents, where the latter is defined as
\begin{eqnarray}
\mathcal{O}^\textrm{con}_\mu(x)=\frac{1}{2}\Big(\overline{\psi}(x+a\hat{\mu})(1+\gamma_\mu)U^\dagger_\mu(x)\psi(x)-\overline{\psi}(x)(1-\gamma_\mu)U_\mu(x)\psi(x+a\hat{\mu})\Big)
\end{eqnarray}
where $\psi=u,d$. In principle, we are able to determine the EM form factors and $\langle x \rangle$ for the proton and for the neutron, 
depending on the linear combination of contributions from the quark 
correlation functions. However, here we focus on the iso-vector combination for which the quark-disconnected diagrams cancel.  
To improve the overlap of the interpolating operators with the nucleon, we use Gaussian smearing \cite{smear:Gaussian89},
supplemented by APE smeared links \cite{Albanese:1987ds}, at both source and sink. 

~\newline
\noindent
The calculation of the three-point function involves the insertion of an operator at time $t$; 
to do this we use the `fixed sink method', which fixes the final and initial states whilst allowing 
both the operator and momentum transfer to be chosen without the need for additional inversions 
\cite{Martinelli:1988rr}. Our specific choice of kinematics $\vec{p}^\prime=0$, and thus $\vec{p}=-\vec{q}$, allow us to 
extract all vector form factors $G_E$, $G_M$, ($F_1$, $F_2$) as well as both the axial charge $g_A$ and  
$\langle x \rangle$ (when $\vec{q}\rightarrow 0$) from eq.~(\ref{Ratio}) at large time arguments, 
\begin{equation}
R_{\gamma_5\gamma_3}(\vec{q}=0,t,t_s)=g_A,
\end{equation}
\begin{equation}
R_{\mathcal{O}_{\langle x \rangle}}(\vec{q}=0,t,t_s)=m_N\langle x \rangle^{\mathrm{bare}},
\end{equation}
\begin{equation}
R_{\gamma_0}(\vec{q},t,t_s)=\sqrt{\frac{M+E}{2E}}G_E(Q^2),
\end{equation}
\begin{equation}
R_{\gamma_i}(\vec{q},t,t_s)=\epsilon_{ij}p_j\sqrt{\frac{1}{2E(E+M)}}G_M(Q^2),\quad i=1,2.\label{R_GM}
\end{equation}
\section{Systematics of extraction\label{seciii}}
\begin{figure}
\centering
\hspace*{-0.45in}
\includegraphics[width=0.60\linewidth]{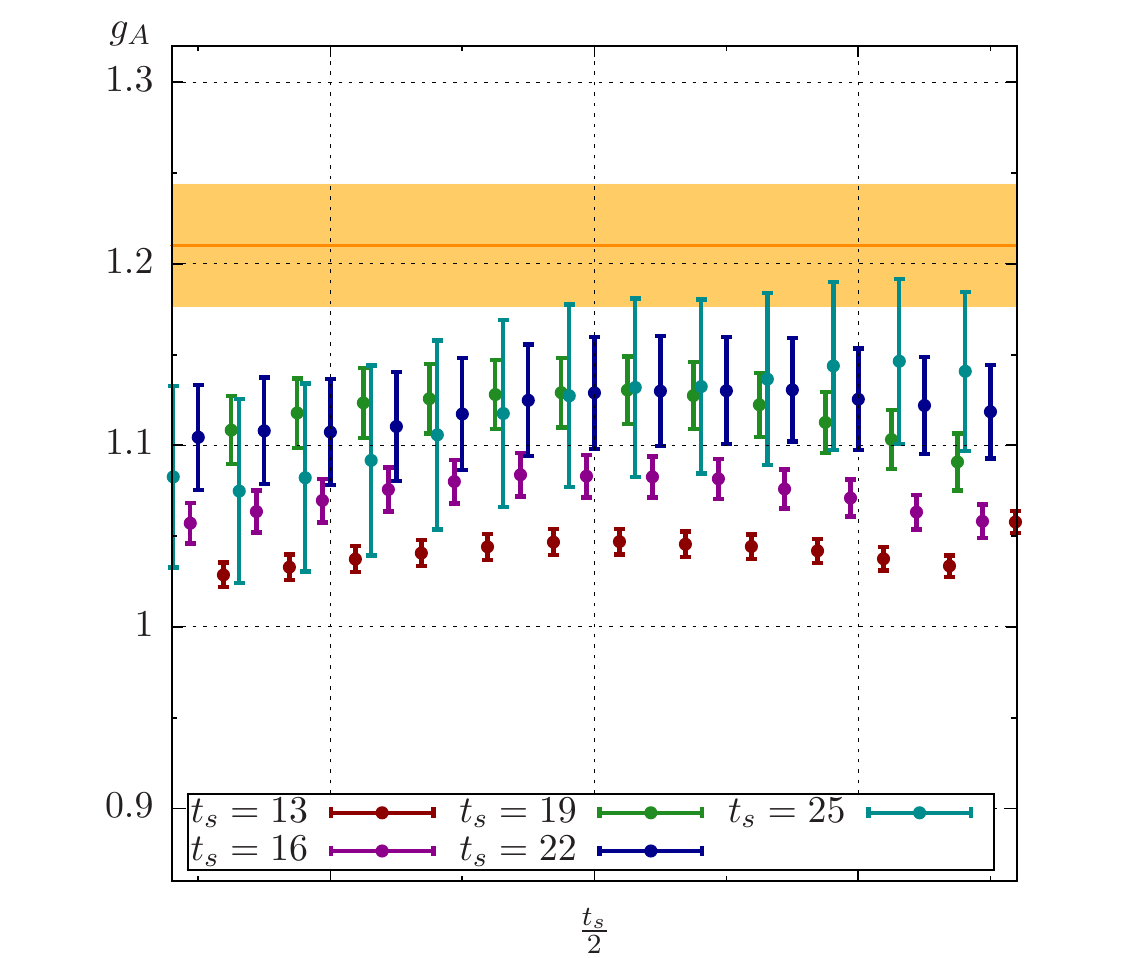}\hspace*{-0.45in}\includegraphics[width=0.60\linewidth]{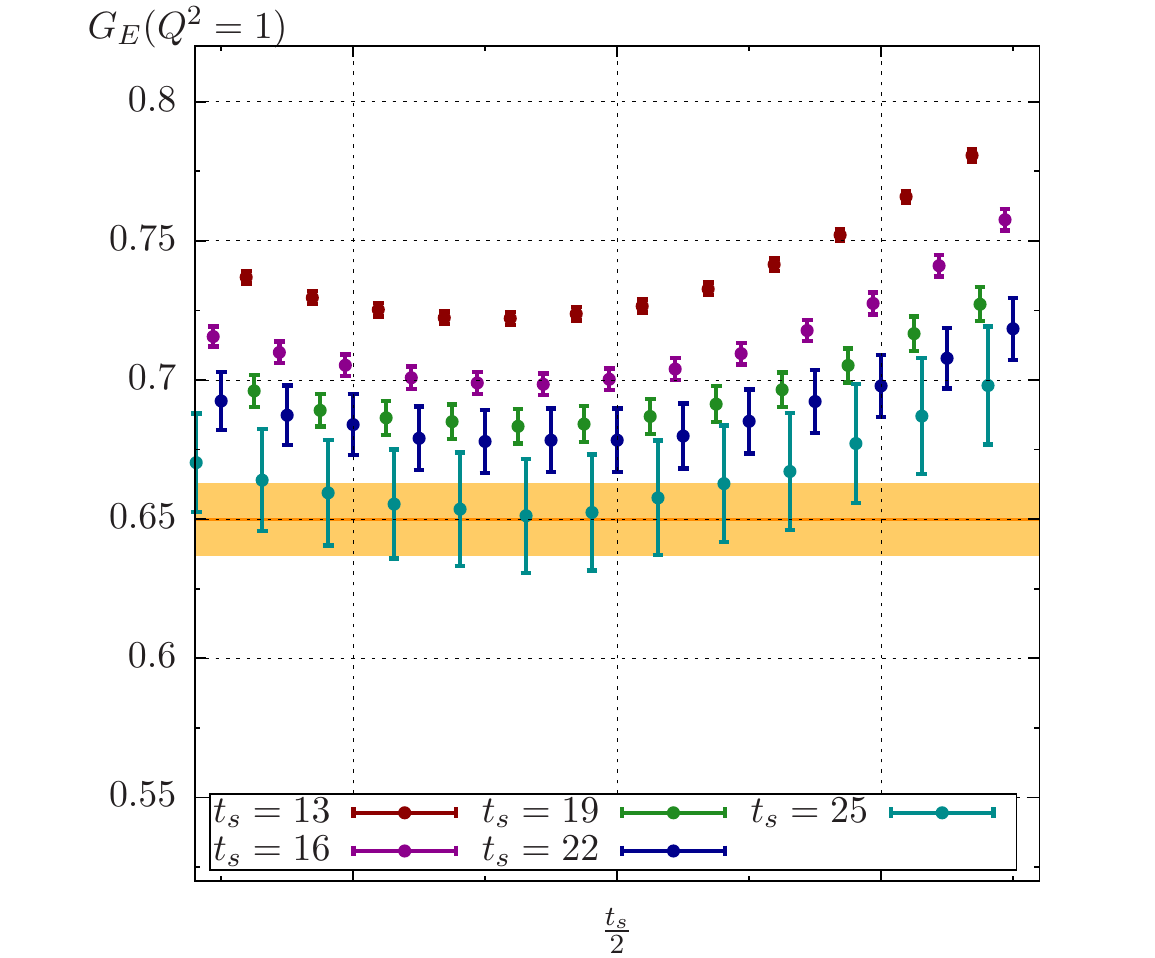}\\
\includegraphics[width=0.60\linewidth]{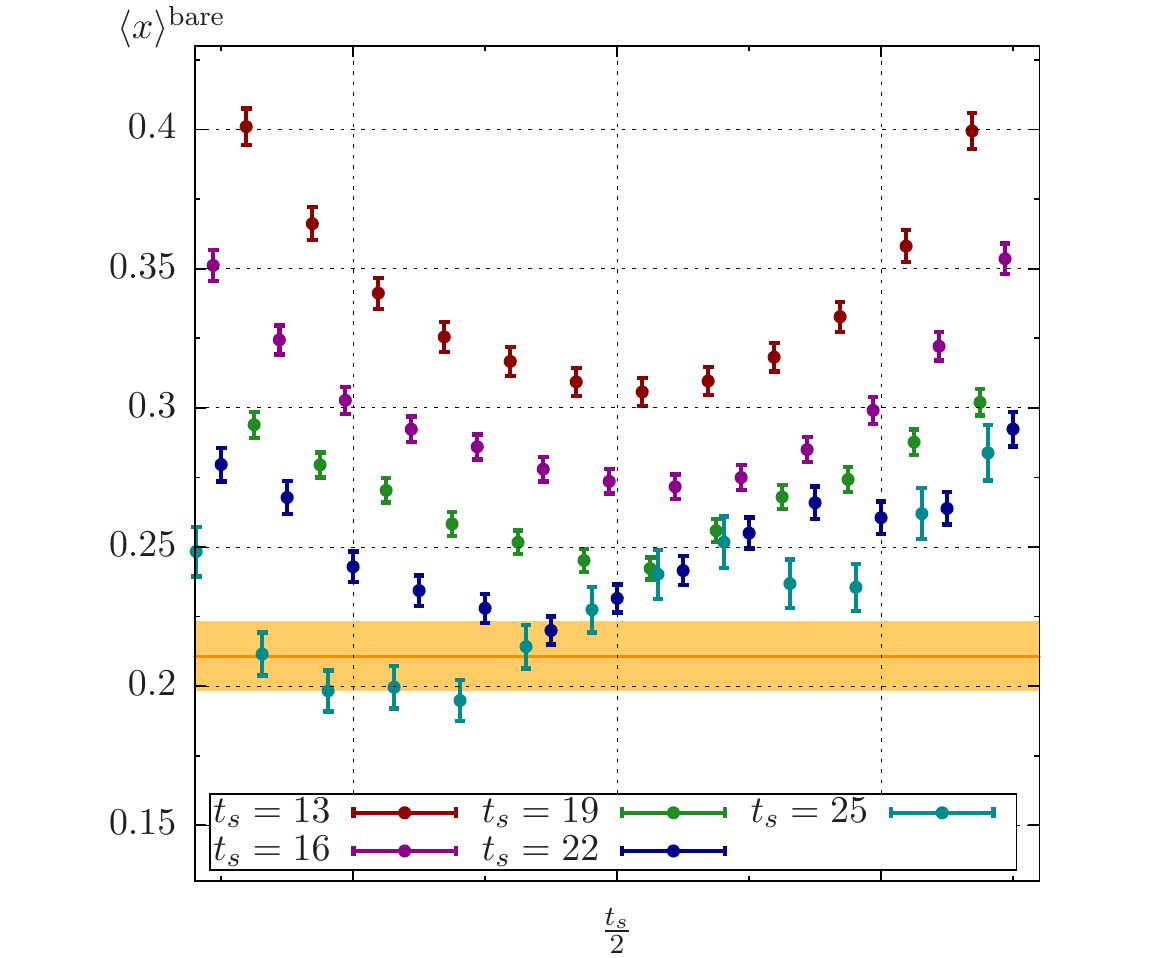}
\caption{\label{ratio_N6} $g_A$, $G_E\big(Q^2=1(\frac{2\pi}{La})^2\big)$ and $\langle x \rangle^{\mathrm{bare}}$ for different source-sink 
separations, $t_s$, indicated by the legend. The summation method result is given by the yellow band. 
All panels are shown for our `N6' ensemble $(m_\pi=340~\textrm{MeV})$. Due to the large statistical noise the 
largest $t_s/a=28$ data has been removed.}
\end{figure}

\noindent
In order to have an unbiased determination of the quantities of interest, the correlation functions must have reached their asymptotic behaviour. 
If the asymptotic behaviour has not been 
reached simple plateau fits will show a systematic trend that is dependent on the source-sink separations $t_s$ (fig.~\ref{ratio_N6}). \emph{A priori} 
it is not possible to know the appropriate source-sink separation for a given quantity, which also depends on the projection properties of the nucleon
interpolating operators. 
~\newline

\begin{figure}
\centering
\hspace*{-0.42in}\includegraphics[width=1.12\linewidth]{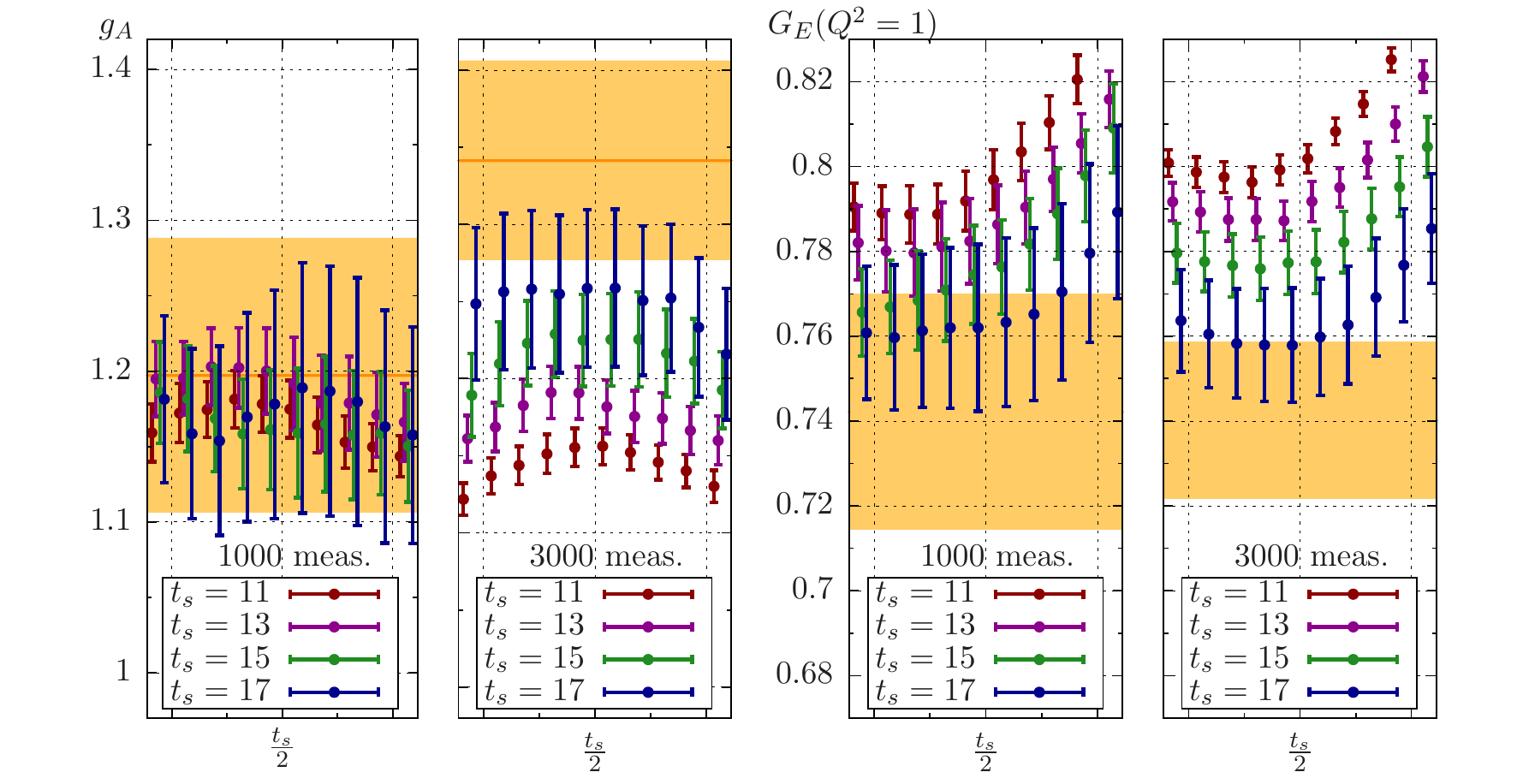}\hspace*{-0.4in}
\caption{\label{ratio_F7} Left panels: $g_A$. Right panels: $G_E\big(Q^2=1(\frac{2\pi}{La})^2\big)$. For each the left sub panel 
shows the quantity for 1000 measurements and the right sub-panel for 3000 measurements. 
Shown for the `F7' ensemble $(m_\pi=277~\textrm{MeV})$.}
\end{figure}

\noindent
For all our ensembles we use four separate source-sink separations ($\sim 0.6 -1.2$ fm), and even at the largest $t_s\sim1.1$ fm it is not
clear from a plateau fit that the contaminations from excited states are sufficiently suppressed for an unbiased determination of the quantities.
Therefore, for the `N6' ensemble we extended the number of source-sink separations from four to six, increasing the source-sink separation range up to $t_s/a=28$ (1.4 fm), see fig.~\ref{ratio_N6}. Even for this extended range, it is difficult to determine that the data has reached the asymptotic behaviour, before the signal is lost for the largest $t_s/a=28$, as can be seen from the systematic trend in the data sets.
It is therefore important to take the excited states into account to have a good handle on possible
systematic errors. The excited-state contributions to the ratio may be factorised from the ground 
state contributions, so that
\begin{eqnarray}
R(\vec{q},t,t_s)=R^0(\vec{q},t,t_s)\Big(1+\mathcal{O}\big(e^{-\Delta t}\big)+\mathcal{O}\big(e^{-\Delta^\prime(t_s-t)}\big)\Big),
\end{eqnarray}
where $\Delta$ and $\Delta^\prime$ are the energy gaps of the initial and final nucleons respectively.  
The method of summed operator insertions \cite{Sum},
\begin{equation}
S(t_s)=\sum_{t=0}^{t_s}R(\vec{q},t,t_s)\rightarrow c(\Delta,\Delta^\prime)+t_s\left(G_{E,M}+\mathcal{O}\big(e^{-\Delta t_s}\big)+\mathcal{O}\big(e^{-\Delta^\prime t_s}\big) \right),
\end{equation}
allows the form factors to be extracted from the slope after computing $S(t_s)$ for several $t_s$. 
The results for the summation method are overlaid in yellow in figs.~\ref{ratio_N6} and~\ref{ratio_F7}
and tend to agree or are minimally overlapping with the data for $t_s/a=25$, corresponding to $t_s=1.25$~fm, 
therefore indicating that the asymptotic behaviour has not yet been reached.
A common method is to fit the largest $t_s$ data with a plateau. However it is difficult, as mentioned,
to know \emph{a priori} if the source-sink separation is `large enough', whereas the summation method has
the advantage that the excited states are parametrically reduced and there is no need to fit a plateau 
to what can sometimes be very noisy data, especially for large $t_s$ and large $Q^2$. Also, the summation 
method only requires linear fits, whereas any extension of plateau fits to include excited states would 
imply non-linear (and therefore possibly unstable) fits.
~\newline

\noindent
Another important consideration besides the source-sink separations, is the question of: 
what is `enough' statistics in order to satisfactorily resolve the desired 
quantity? To check this, on the `F7' ensemble $(m_\pi=277~\textrm{MeV})$, we show $g_A$ 
and $G_E\big(Q^2=(\frac{2\pi}{La})^2\big)$ with both 1000 and 3000 measurements (fig.~\ref{ratio_F7}). The results for 1000 
measurements suggest that the largest $t_s$ for the plateau method is `large enough' as the
$t_s$ dependence appears to have saturated. However, 
when the statistics are increased to 3000 measurements we clearly see that this is not the case,  
as indicated by the reduced overlap between the summation method and the individual $t_s$ data sets in both quantities. 

\section{Electromagnetic form factor $Q^2$ dependence\label{seciiii}}
\noindent
For the discussion of the electromagnetic form factors we concentrate on the conserved current 
as this removes the requirement of any renormalisation for the lattice operators; however, we note 
that a comparison between the local and conserved current provides a check of the renormalisation 
factor, which we find to be in agreement with other work (such as~\cite{Della Morte:2005rd}).
\begin{figure}
\centering
\includegraphics[width=0.80\linewidth]{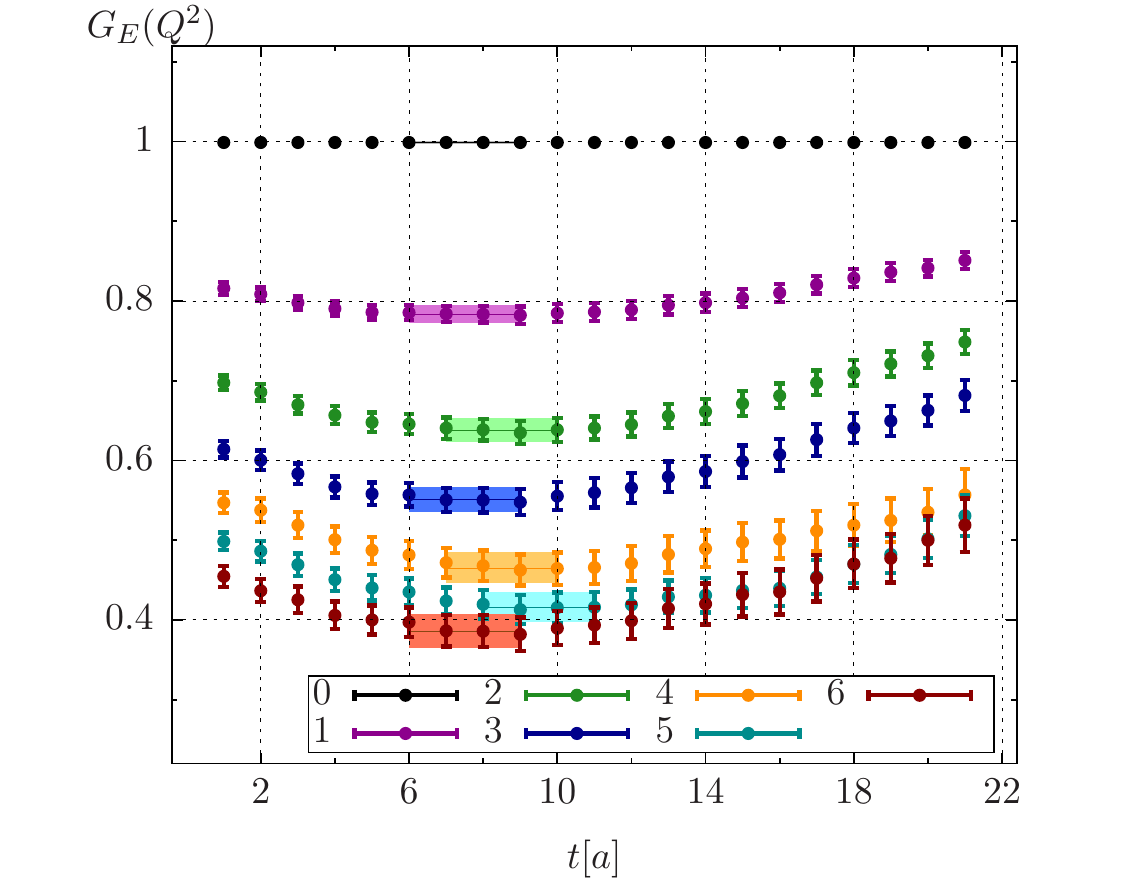}
\caption{\label{GE_qdep} $G_E$ for each $Q^2=n(\frac{2\pi}{La})^2$ where the legend gives the value of $n$. Shown for the largest $t_s\sim1.1$~fm, for the `O7' ensemble $m_\pi=270~\textrm{MeV}$.}
\end{figure}
~\newline

\begin{figure}
\centering
\hspace*{-0.45in}\includegraphics[width=0.60\linewidth]{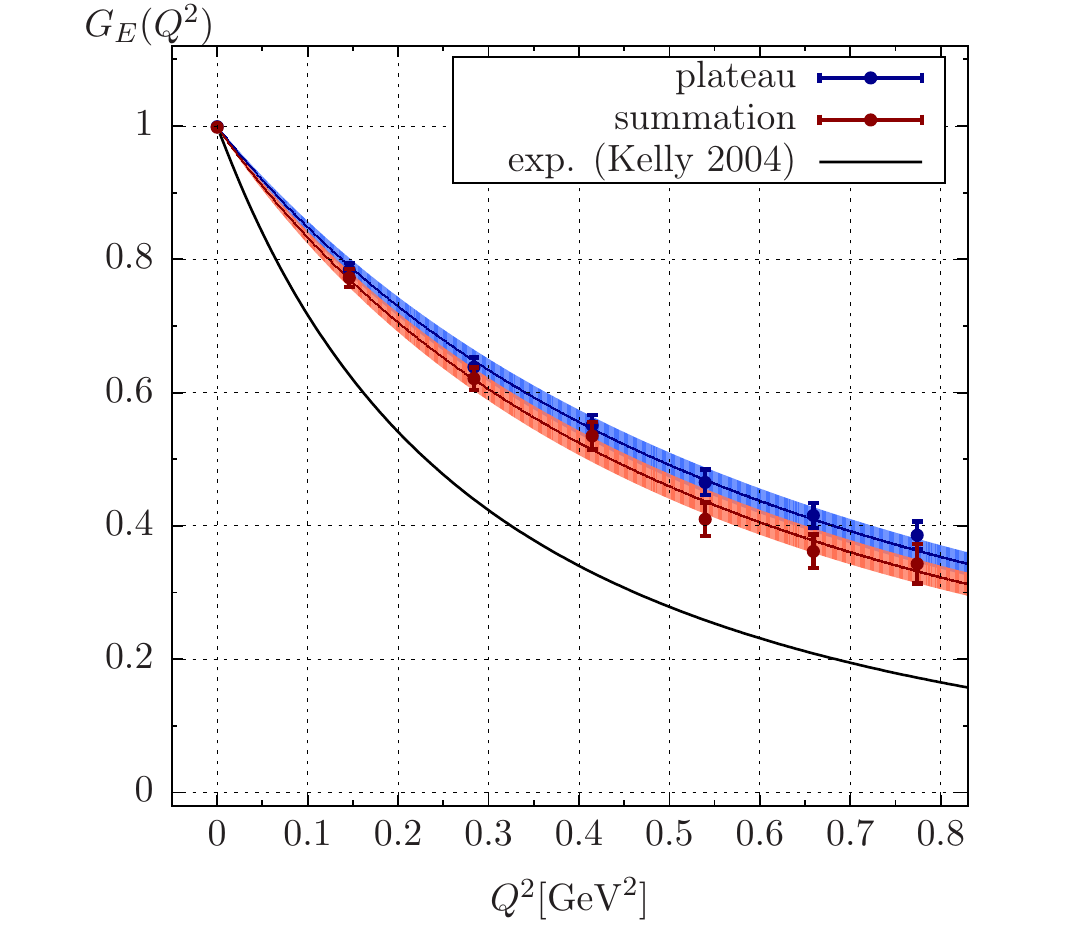}\hspace*{-0.45in}\includegraphics[width=0.60\linewidth]{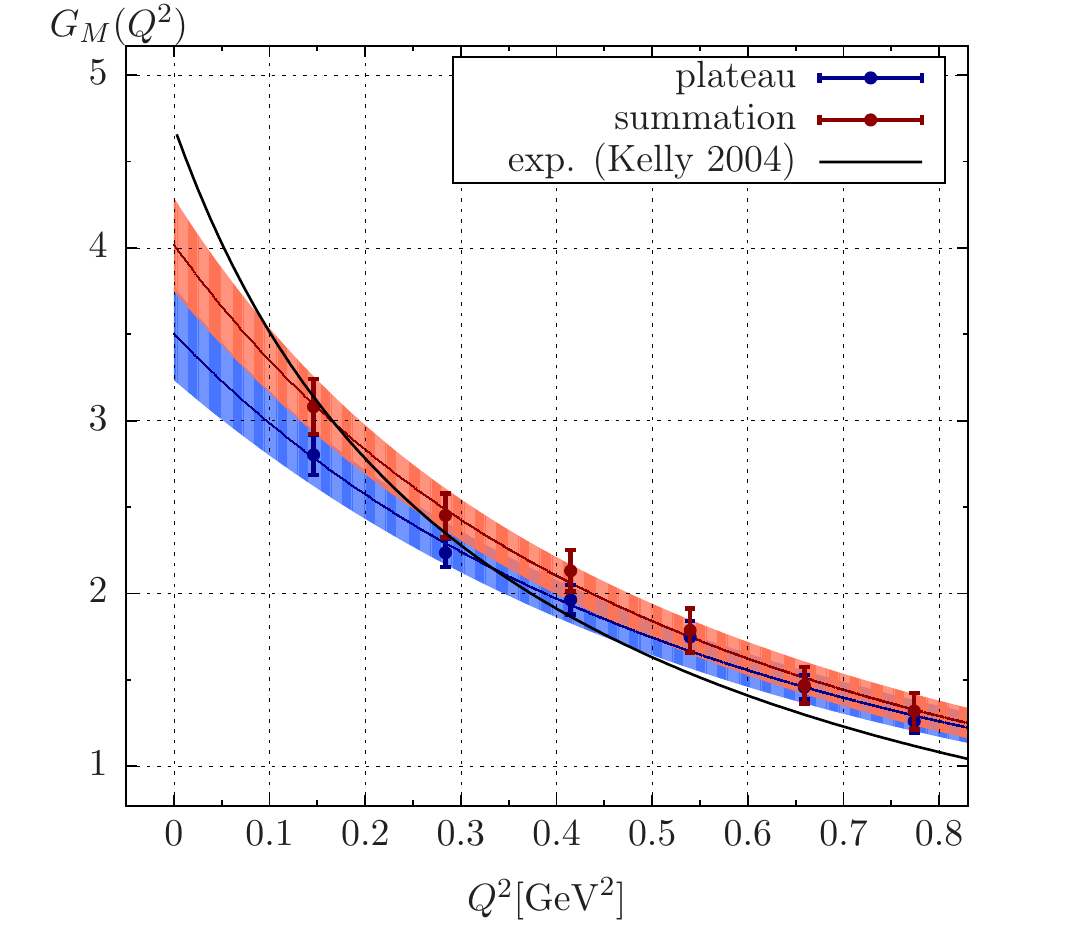}
\includegraphics[width=0.60\linewidth]{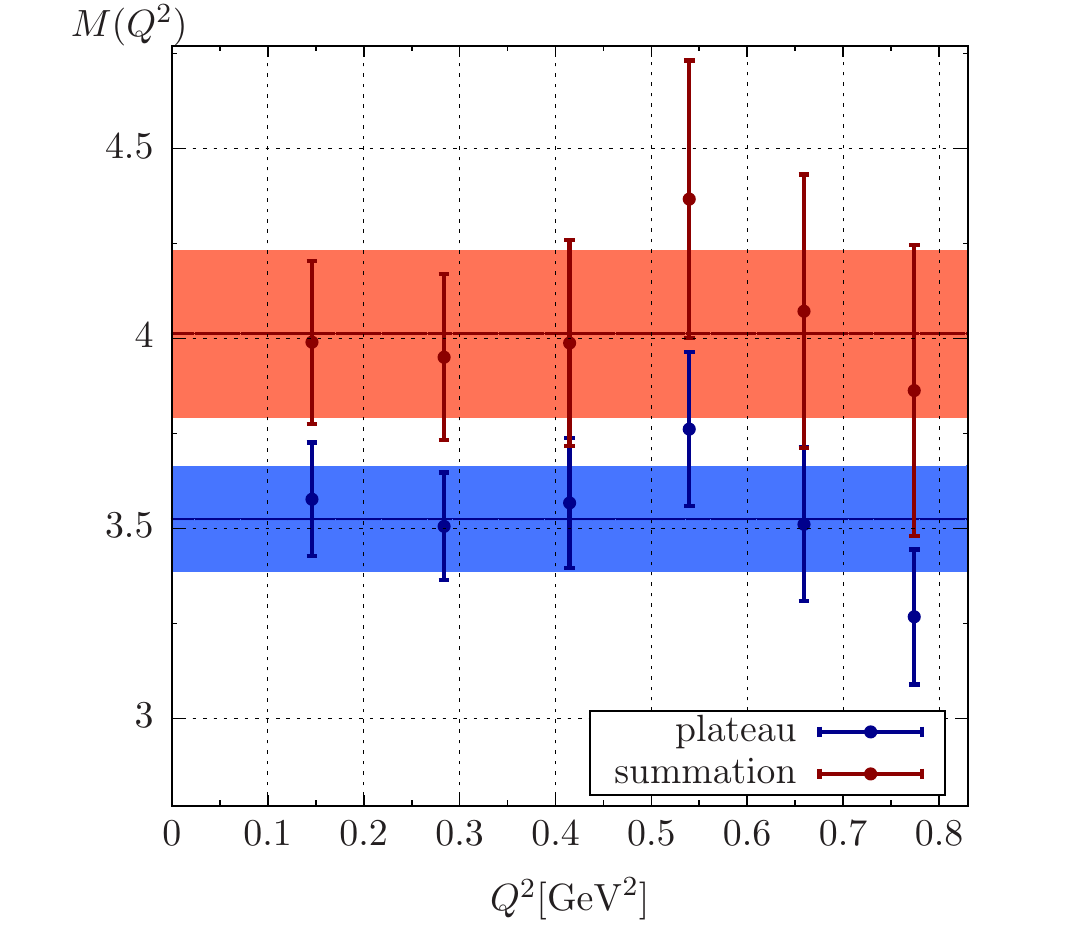}
\caption{\label{qdep} The top left and right panels show the $Q^2$ dependence of $G_E$ and $G_M$ respectively, which may be compared to the Kelly parameterisation \cite{Kelly:2004hm} of the experimental data.
The bottom panel shows $M(Q^2)$. The lattice data is for the 
`O7' ensemble $m_\pi=270~\textrm{MeV}$.}
\end{figure}

\noindent
To model the $Q^2$ dependence of the form factors, shown for $G_E$ in fig.~\ref{GE_qdep},  we use
a dipole ansatz
\begin{equation}
G_{E,M}(Q^2)=G_{E,M}(0) \big/\left(1+Q^2/M_{E,M}^2\right)^2,\label{dipole}
\end{equation}
shown in fig.~\ref{qdep} (for the `O7' ensemble, $m_\pi=270~\textrm{MeV}$) for $G_E$ and $G_M$ alongside 
the Kelly parameterisation \cite{Kelly:2004hm} of the experimental data. It should be noted that in order for the lattice data 
and experimental parameterisation to be fully compatible a chiral extrapolation of the lattice data is required. 
In the case of $G_E$ we see a better agreement with the Kelly
parameterisation \cite{Kelly:2004hm} for the summation method than for a plateau fit at the largest $t_s\sim1.1$~fm, especially at large $Q^2$.
However, for the case of $G_M$ it is harder to disentangle the plateau and summation 
methods, and further study is required to determine whether or not this is indicative that the asymptotic behaviour 
has been reached, or if still more statistics are required. Due to the extra momentum factor required in the extraction of $G_M$
(see eq.~\ref{R_GM}), its statistical accuracy is worse than that of $G_E$.
The charge radii can be extracted from the dipole mass. However, due to the absence of a measured point at 
$G_M(Q^2=0)$, the determination of the radius, which is effectively the slope of the form factor at $Q^2=0$, 
is less constrained for $\langle r_M^2 \rangle$ than $\langle r_E^2 \rangle$, as can also be seen in fig.~\ref{qdep}.
~\newline

\noindent
We may obtain the magnetic moment $\mu$ from $G_M(Q^2=0)$ and also from the ratio
\begin{equation}
M(Q^2)=\frac{G_M(Q^2)}{G_E(Q^2)},\quad\mathrm{where}\quad\mu=M(0)=1+\kappa
\end{equation}
shown in fig.~\ref{qdep}. The effect of this ratio is to cancel the $Q^2$ behaviour, indicating that the form factors $G_E$ and $G_M$ have a 
very similar shape and hence their radii are quite similar. We can therefore extract $\mu$ from a constant fit to the data, which is 
compatible within errors to $G_M(Q^2=0)$ for both the summation and plateau methods. 

\section{Chiral dependence of the EM form factors and axial charge\label{seciiii}}
\begin{figure}
\centering
\includegraphics[width=0.8\linewidth]{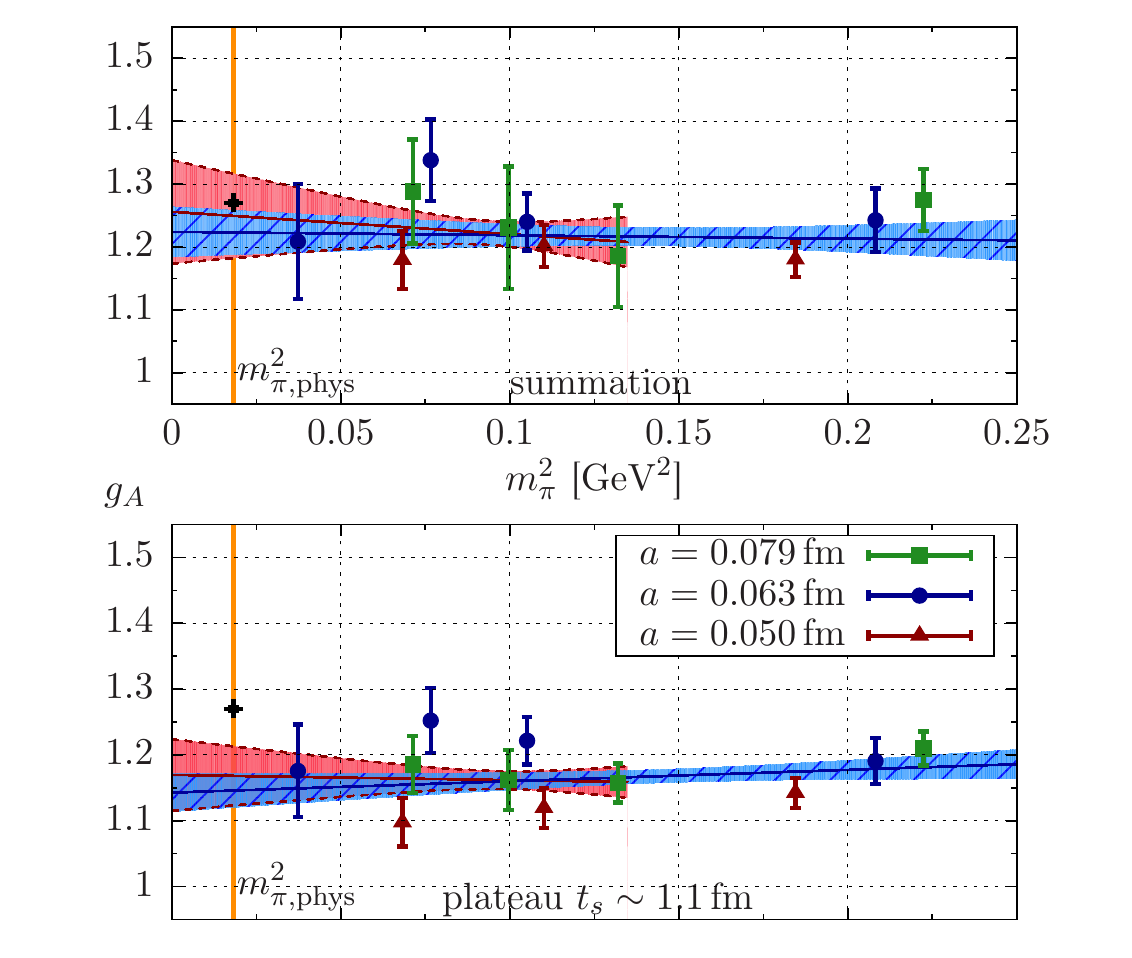}
\caption{\label{chi_gA} Chiral extrapolation of $g_A$ to the physical point (vertical yellow line). 
The black cross shows the experimental result \cite{Beringer} and the different symbols indicate the lattice spacing 
(see legend). The blue band shows a linear fit to the entire range, whereas the red band shows a 
linear fit with a mass cut at $m_\pi=360$~MeV.}
\end{figure}

\begin{figure}
\centering
\hspace*{-0.30in}\includegraphics[width=1.15\linewidth]{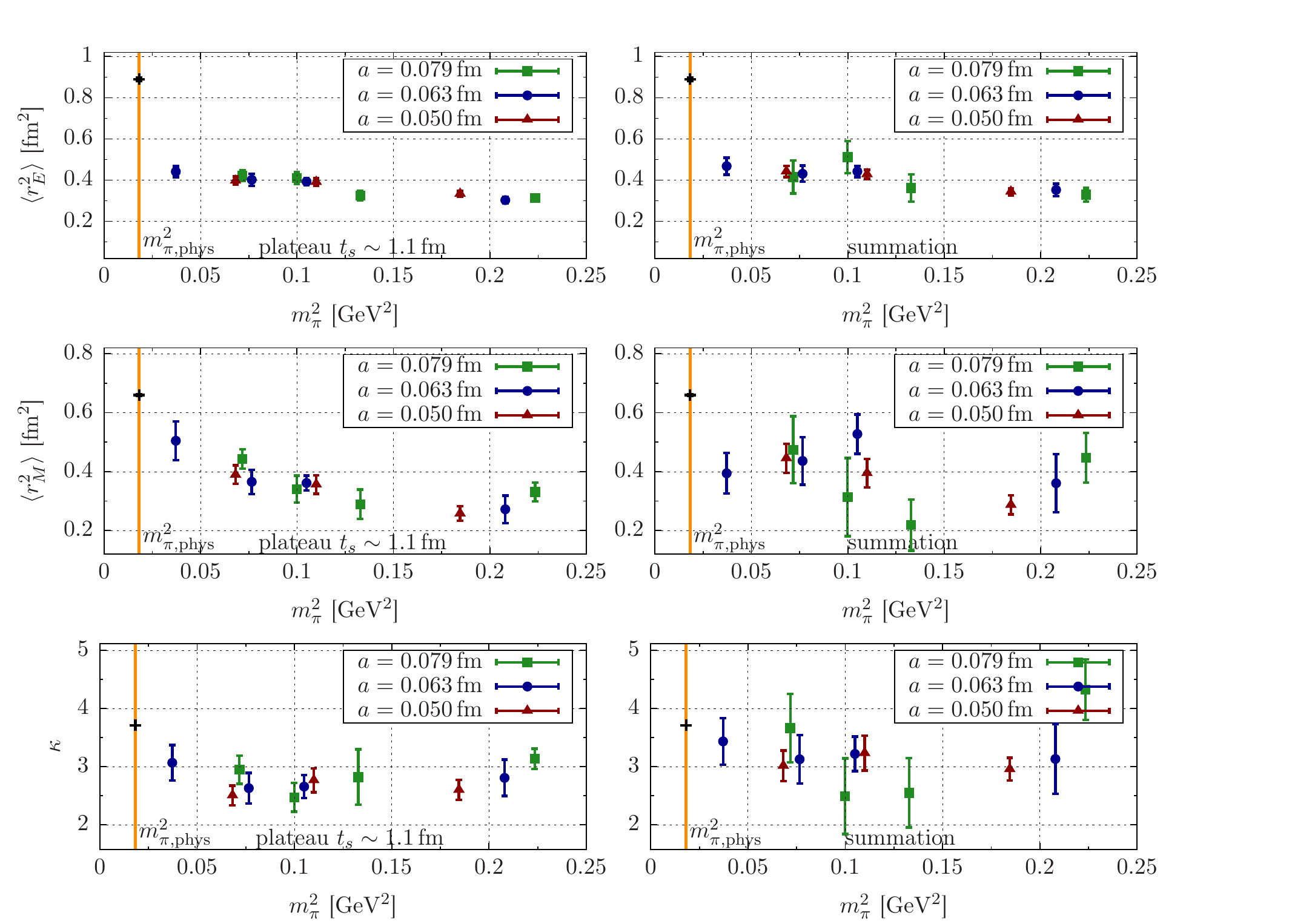}
\caption{\label{chi} Chiral dependence of $\langle r_E^2\rangle$, $\langle r_M^2\rangle$ and $\kappa$. 
The vertical yellow line shows the physical point and the black cross the experimental result \cite{Beringer}. The different symbols indicate the lattice spacing (see legend). }
\end{figure}

\noindent
The ensembles listed in table~\ref{ensembles} cover a range of pion masses, from 195 to 650~MeV, 
enabling us to both perform extrapolations in the pion mass to the physical points and to check finite-volume 
and discretisation effects for all quantities. All of the chiral dependence plots figs.~\ref{chi_gA} and \ref{chi} 
show the data for different lattice spacings in different colours, given in the legend. The experimental value 
is shown by a black cross at the physical point (yellow vertical line).
~\newline

\noindent
The individual data points in fig.~\ref{chi_gA} exhibit only a mild $m_\pi$ dependence, 
and so a linear fit of the form
\begin{equation}
 A+Bm_\pi^2,
\end{equation}
may be appropriate to model the $g_A$ data. To check the stability of the chiral extrapolation
to the entire pion mass range, 
we have applied a cut at $m_\pi=360$~MeV, for which we see that the two extrapolations agree 
very well within statistical precision. In addition we see no obvious finite-size or discretisation effects; 
the latter has been checked with the addition of an $a^2$ term to the fits. 
If the excited states are taken into account via the summation method, we obtain a 
value for $g_A$ that is compatible with the experimental result. 
By contrast, using the plateau method with a source-sink separation of 
$\sim1.1$ fm yields discrepancy with the experimental result \cite{Beringer}, which is larger than $1\sigma$,
regardless of the pion mass range used for the extrapolation.
Therefore, the summation method provides strong evidence that excited states need to be sufficiently accounted for to reach 
agreement with experimental values
~\newline

\noindent
The chiral dependence for $\langle r_E^2\rangle$, $\langle r_M^2\rangle$ and $\kappa$ are shown in fig.~\ref{chi}.  
With the exception of $\langle r_M^2\rangle$, the comparison of the plateau and summation method indicates that, 
as for $g_A$, it is necessary to account for excited states. 
For $\langle r_M^2\rangle$, we see that any extrapolations to the physical point will be
strongly dependent upon the most chiral point and we note that larger statistical
errors and fluctuations within the $\langle r_M^2\rangle$ data are largely due to 
the absence of a point equivalent to $G_E(Q^2=0)=1$, which helps to constrain both the $Q^2$ behaviour
and the determination of the charge radius. As for $g_A$, the EM form factor data also shows 
no obvious finite volume or discretisation effects. We are currently exploring the effect 
of other fit forms on the results in figs.~\ref{chi_gA} and \ref{chi}, including ans\"atze based
on HBChPT, so as to have a comprehensive picture of the systematic effects. This will be commented upon in a 
forthcoming paper \cite{FFpaper}.

\section{Conclusions and outlook\label{seciiiii}}
\noindent
We have presented preliminary results for the nucleon's axial charge, vector form factors and quark 
momentum fraction with a focus on systematic errors due to excited state contaminations. 
Chiral extrapolations for the axial charge show that the use of the summed insertions method 
effectively accounts for contaminations from excited states, leading to good agreement with experiment.
However, whilst we note that we are still analysing the chiral behaviour for this and the electromagnetic form factors, 
our data indicates that with sufficient statistics excited-state effects can be resolved and we present
evidence that the summation method is an important tool to control the associated systematic errors. 
Further to this, we see no obvious finite size or discretisation effects in our data for all quantities.
~\newline

\noindent
For the average quark momentum fraction $\langle x \rangle$, which we have, so far, only evaluated
at the bare ratio level, we see a similar effect as is seen for the EM form factors and axial charge, and 
conclude that excited-state effects appear to be equally important here, and reliably controlling them
could help improve agreement with experiment as has proven to be the case for most of the other 
quantities discussed here. The average quark momentum fraction 
will be fully analysed including chiral fits, once a calculation of the required renormalisation 
constants using a non-perturbative scheme is completed.
~\newline

\noindent
The axial charge and form factors analysis is in the process of being finalised and will appear in 
an upcoming publication \cite{FFpaper}, therefore all results in this proceedings should, at the moment, be considered
preliminary.

\section*{Acknowledgments}
\noindent
Our calculations were performed on the ``Wilson'' HPC Cluster at the Institute for Nuclear Physics, University of Mainz. We thank 
Christian Seiwerth for technical support. We are grateful for computer time allocated to project HMZ21 on the BG/Q ``JUQUEEN'' computer at NIC, J\"ulich. This work was granted access to the HPC resources of the Gauss Center for Supercomputing at Forschungzentrum J\"ulich, Germany, made available within the Distributed European Computing Initiative by the PRACE-2IP, receiving funding from the European Community's Seventh Framework Programme (FP7/2007-2013) under grant agreement RI-283493. This work was supported by the DFG via SFB 1044 and grant HA 4470/3-1. We are grateful to our colleagues within the CLS initiative for sharing ensembles.

\end{document}